\documentclass[onecolumn,draftclsnofoot,10pt]{IEEEtran}

\usepackage{epstopdf}
\usepackage[draft]{hyperref} 
\usepackage{setspace}
\usepackage{amsmath}
\usepackage{amssymb}
\usepackage{amsthm}
\usepackage{amsfonts}
\usepackage{color}
\usepackage{graphicx}
\usepackage{subfigure}  
\usepackage{caption}
\usepackage{pdfpages}
\usepackage{cite}
\usepackage{mathtools}
\usepackage{stmaryrd}
\usepackage{xcolor}
\newtheorem{theorem}{Theorem}
\usepackage[]{algorithmicx}
\usepackage{algpseudocode,algorithm}

\usepackage[mathscr]{euscript}
\usepackage{mathrsfs}

\usepackage{soul}
\newtheorem{lemma}{Lemma}

\def\@fnsymbol#1{\ensuremath{\ifcase#1\or *\or \dagger\or \ddagger\or
   \mathsection\or \mathparagraph\or \|\or **\or \dagger\dagger
   \or \ddagger\ddagger \else\@ctrerr\fi}}

\begin{document}
\title{Unlabelled Sensing with Priors: Algorithm and Bounds}
%

\author{Garweet Sresth$^{\dagger}$ \qquad Ajit Rajwade$^{\star}$ \qquad Satish Mulleti$^{\ddag}$

\thanks{Garweet Sresth is with the Centre for Machine Intelligence and Data Science, Indian Institute of Technology Bombay,  India. Ajit Rajwade is with the Department of Computer Science and Engineering, Indian Institute of Technology Bombay, India. Satish Mulleti Department of Electrical Engineering, Indian Institute of Technology Bombay, Mumbai, India. \\
Emails: garweetsresth@gmail.com, ajitvr@cse.iitb.ac.in, mulleti.satish@gmail.com}
}

\maketitle
\begin{abstract}
In this study, we consider a variant of unlabelled sensing where the measurements are sparsely permuted, and additionally, a few correspondences are known. We present an estimator to solve for the unknown vector. We derive a theoretical upper bound on the $\ell_2$ reconstruction error of the unknown vector. Through numerical experiments, we demonstrate that the additional known correspondences result in a significant improvement in the reconstruction error. Additionally, we compare our estimator with the classical robust regression estimator and we find that our method outperforms it on the normalized reconstruction error metric by up to $20\%$ in the high permutation regimes $(>30\%)$. Lastly, we showcase the practical utility of our framework on a non-rigid motion estimation problem. We show that using a few manually annotated points along point pairs with the key-point (SIFT-based) descriptor pairs with unknown or incorrectly known correspondences can improve motion estimation.  
\end{abstract}


\section{Introduction}
Estimating an unknown vector from a set of linear and noisy measurements is a well-known problem in many applications that can be solved using least squares. In this scenario, the measurements and the unknown vector are connected by a measurement matrix. Usually, it is presumed that each measurement relates to a specific row in the matrix. Nevertheless, due to vagaries in the measurement process, the correspondence between the rows and the measurements, whether in part or entirely, might be lost. The objective now becomes estimating the unknown vector from an unknown permutation of the measurements.

The aforementioned problem is known as unlabelled sensing, and it arises naturally in many engineering and biological applications. For example, in the point-matching problem\cite{pointmatching}, the objective is to determine the unknown underlying permutation given the correspondences between the two point sets. Other areas where the unlabelled sensing problem and its variations naturally arise are group testing \cite{dorfman,covidgrouptesting}, record-linkage \cite{recordlinkage, shi2019spherical}, simultaneous pose and correspondence determination in computer vision \cite{softposit, subspacematching}, simultaneous localization and mapping (SLAM) in robotics \cite{slam}, data de-anonymization in security and privacy \cite{datadeanonymization, privacyinmedicaldata}, and data collection in sensor networks \cite{datacollection}.

Estimation of the unknown vector from their linear permuted and noisy measurements is, generally, an ill-posed problem unless constraints are imposed on the permutation level, signal-to-noise ratio (SNR), and the number of measurements. For example, Unnikrishnan et al. \cite{vetterli} showed that a $d$ dimensional vector can be uniquely identified with high probability from its $N$ linear permuted measurements without noise iff $N\geq 2d$. With an additional assumption that the unknown vectors are generic, $N>d$ measurements are shown to be sufficient in the absence of noise \cite{tsakiris2020algebraic}. On the other hand, in \cite{pananjady_2018}, thresholds on SNR are established for reconstruction. These theoretical guarantees are independent of any algorithm. 

On the algorithm front, the problem of estimation is commonly framed as a least-squares challenge or as a maximum likelihood estimation problem. The goal is to minimize the objective function for both the unidentified vector and the permutations. However, as the optimization must encompass all possible $N!$ permutations, any algorithm becomes computationally impractical without extra limitations such as the ordering of the unknown correspondence \cite{haghighatshoar2017us}, multiple measurements with the same permutation matrix  \cite{pananjady2017denoising}, and low-dimensional unknown vectors, \cite{abid2017linear, elhami_algo_2017,hsu2017linear,tsakiris2019homomorphic,tsakiris2020algebraic,peng2020linear}. In addition, a few works exploit the sparsity of the unknown vectors to make the problem less challenging (see \cite{peng2021homomorphic_sparse} and references therein).

The aforementioned algorithms assume the possibility of losing complete correspondence among the measurements. In contrast, \cite{bendavid} considers a scenario where only a few correspondences are lost. The assumption is valid in many applications. For example, in point-matching, employing high-accuracy algorithms ensures that only a few correspondences are wrongly matched between the two sets. With few lost correspondences or equivalently assuming sparse permutations, the error due to permutation can be treated as sparse outliers \cite{bendavid}. While the estimation of unknown vectors in the presence of these sparse outliers can be accomplished using robust regression techniques (refer to \cite{robustregression}), \cite{bendavid} suggests an alternative approach that is robust to noise and employs $\ell_1$ regularization. The authors have established upper bounds on the errors associated with estimating the unidentified vector and permutations. Through simulations, they demonstrated that their algorithm can successfully recover vectors for up to $50\%$ of the permutations.

Apart from the requirement for sparse permutations, many situations allow for obtaining a limited number of measurements with correct correspondences. For instance, in the point-matching problem, it is possible to manually annotate a small set of precise point pairs with the help of a subject matter expert. This leads to the question: How can this additional information be optimally incorporated while upholding the assumption of sparse permutations?

In this paper, we consider an adaptation of the standard unlabelled sensing problem, where we assume that the measurements are sparsely permuted, and a few measurements with correct correspondences are available. Within these settings, inspired from \cite{bendavid}, we have formulated an $\ell_1$-regularized problem and derived an upper bound on the estimation error of the unknown vector in terms of noise variance, dimension $d$, number of measurements $N$, sparsity level of the permutation matrix, and number of correct correspondences. We show how the estimation error falls down as the number of known correspondences increases. We solved the problem using an off-the-shelf solver and compared our approach with robust regression. Our method outperforms the robust regression method on the normalized reconstruction error metric by up to $20\%$  in the high permutation regimes $(>30\%)$. We show that a few known correspondences can significantly reduce the reconstruction error compared to a scenario without knowledge of any correct correspondence. As an application, we consider the point-matching problem. We show that using a few manually annotated point pairs results in a visually better reconstruction.

The organization of the paper is as follows. In Section 2, we formally define the measurement model and the optimization problem. Theoretical bounds are presented in Section~3, whereas Section~4 discusses the numerical analysis. In Section~5, we show results for the point-matching problem followed by conclusions.

\section{Problem Formulation}
Consider a set of linear measurements $\boldsymbol{A}\boldsymbol{x}_0$ where $\boldsymbol{x}_0 \in \mathbb{R}^d$ is the unknown vector to be estimated and $\boldsymbol{A} \in \mathbb{R}^{N \times d}$ with $N\geq d$ is the sensing matrix. In general, the problem can be solved using least squares, even in the presence of noise. However, in many applications, the measurements are permuted, as discussed in the previous section. In this case, the measurements in the presence of noise are given as
\begin{equation}\label{obs_model}
    \boldsymbol{y} = \boldsymbol{P}_0\boldsymbol{A} \boldsymbol{x}_0 + \boldsymbol{\epsilon},
\end{equation}
where $\boldsymbol{\epsilon}$ is the noise term and $\boldsymbol{P}_0$ is an $N\times N$ permutation matrix. The problem of estimating $\boldsymbol{x}_0$ from $\boldsymbol{y}$ is ill-posed for $N=d$ even if $\boldsymbol{\epsilon} = \boldsymbol{0}$ and $\boldsymbol{A}$ is invertible. Specifically, for any arbitrary permutation matrix $\boldsymbol{P} (\neq \boldsymbol{P}_0)$ and the vector $\boldsymbol{x} = \boldsymbol{A}^{-1} \boldsymbol{P}^{-1} \boldsymbol{P}_0 \boldsymbol{A}\boldsymbol{x}_0$, we have that $\boldsymbol{y} = \boldsymbol{PAx}$. Hence, the solution is not unique. Hence, the condition $N>d$ is necessary to solve the problem. However, $N>d$ need not be sufficient, especially in the presence of noise, unless additional assumptions on $\boldsymbol{P}_0, \boldsymbol{A}$, and $\boldsymbol{\epsilon}$. In this work, we make the following assumptions.
\begin{enumerate}
    \item[(A1)] The entries  of $\boldsymbol{A}$ are independent and identically distributed (i.i.d.)  zero mean, unit variance Gaussian random variables.
    \item[(A2)] The entries of noise vector $\boldsymbol{\epsilon}$ are i.i.d. Gaussian random variables with zero mean and variance $\sigma^2$.
    \item[(A3)] Any $m$ correct correspondences of $\boldsymbol{Ax}$ are given where $m < d$.
    \item[(A4)] For the remaining $p = N-m >d$ measurements without correspondences, at max $k$ entries are permuted. 
\end{enumerate}
The randomness assumptions (A1) and (A2) will be useful in deriving theoretical bounds on the estimation accuracy. The assumption of the correct correspondence and sparse permutations in (A3) and (A4), respectively, are inspired by practical scenarios and will be helpful in improving the estimation accuracy. The assumptions $m<d$ and $N-m>d$ are explained by dividing the measurements $\boldsymbol{y}$ into the following disjoint sets of measurements:
\begin{align}
    \boldsymbol{y}_1 = \boldsymbol{A}_1 \boldsymbol{x}_0 + \boldsymbol{\epsilon}_1, \label{eq:y1}\\
     \boldsymbol{y}_2 = \boldsymbol{P}_2\boldsymbol{A}_2 \boldsymbol{x}_0 + \boldsymbol{\epsilon}_2, \label{eq:y2}
\end{align}
where $\boldsymbol{y}_1\in \mathbb{R}^m$ denotes the measurements with correct correspondences and $\boldsymbol{y}_2 \in \mathbb{R}^{p}$ are the remaining measurements. Note that $\boldsymbol{P}_2$ is a $p \times p$ permutation matrix with $k$ permutations. 

From \eqref{eq:y1}, we observe that $\boldsymbol{x}_0$ can be estimated from $\boldsymbol{y}_1$ provided that $m\geq d$ and we do not require $\boldsymbol{y}_2$. To avoid such a trivial scenario, we assume that $m<d$. Next, if we consider a problem of estimating $\boldsymbol{x}_0$ from $\boldsymbol{y}_2$ only, then as discussed earlier, $p> d$ is necessary. The necessity of the assumption $p> d$ may be questionable in the presence of correct correspondences $\boldsymbol{y}_1$. However, we maintain $p>d$ to include the standard sparse unlabelled sensing problem when $m=0$ \cite{bendavid}.

Our objective is to estimate $\boldsymbol{x}_0$ from $\boldsymbol{y}$ under assumptions (A1)-(A4). To this end, we define the permutation error vector as
\begin{align}
    \boldsymbol{z}_0 = \boldsymbol{P}_2 \boldsymbol{A}_2 \boldsymbol{x}_0 - \boldsymbol{A}_2 \boldsymbol{x}_0. \label{eq:z0}
\end{align}
By using $\boldsymbol{z}_0$, we re-write $\boldsymbol{y}_2$ as
\begin{align}
    \boldsymbol{y}_2 = \boldsymbol{A}_2\boldsymbol{x}_0 + \boldsymbol{z}_0 + \boldsymbol{\epsilon}_2. \label{eq:y2_z}
\end{align}
This representation captures the effect of unknown permutation as an additional unknown but signal-dependent term. By combining \eqref{eq:y1} and \eqref{eq:y2_z}, we have that $\boldsymbol{y} = \boldsymbol{A} \boldsymbol{x}_0 + [\boldsymbol{0}_m^{\mathrm{T}} \,\, \boldsymbol{z}_0^{\mathrm{T}} ]^{\mathrm{T}} + \boldsymbol{\epsilon}$, where $\boldsymbol{0}_m$ is an all-zero vector of length $m$ and the superscript $\mathrm{T}$ denotes transpose operation. Since $\boldsymbol{P}_2$ is a $k$-sparse permutation matrix, we have that $\|\boldsymbol{z}_0\|_0 \leq k$. Further, from \eqref{eq:y2_z}, it can be verified that $\|\boldsymbol{z}_0\|_{\infty} \leq 2\|\boldsymbol{Ax}_0\|_{\infty}$. Hence, the term $[\boldsymbol{0}_m^{\mathrm{T}} \,\, \boldsymbol{z}_0^{\mathrm{T}} ]^{\mathrm{T}}$ could be treated as a sparse outlier, and $\boldsymbol{x}_0$ can be estimated by solving robust-regression problem as
\begin{equation}\label{robust_regression_estimator}
    \boldsymbol{\Tilde{x}_{RR}} = \underset{\boldsymbol{x} \in \mathbb{R}^d} {\operatorname{arg} \min} \|\boldsymbol{y-A x}\|_1.
\end{equation}
However, the formulation and the solution ignore the dependency of $\boldsymbol{z}_0$ on $\boldsymbol{x}_0$ and its sparsity. Moreover, it does not make use of the known correspondences.

To address the aforementioned limitations of robust regression, we propose an alternative formulation as  
\begin{equation}\label{eq:opt}
   \underset{\boldsymbol{x} \in \mathbb{R}^d, \boldsymbol{z} \in \mathbb{R}^p} {\operatorname{arg} \min} \|\boldsymbol{y_1-A_1 x}\|_2^2+\|\boldsymbol{y_2-A_2 x-z}\|_2^2 + \lambda \|\boldsymbol{z}\|_1.
\end{equation}
where the first two terms are data-fidelity terms and $\|\boldsymbol{z}\|_1$ is  a sparsity promoting term with a regularization parameter $\lambda$. This problem is convex and can be solved by using an off-the-shelf solver such as CVPXY \cite{diamond2016cvxpy, agrawal2018rewriting}. Let $\boldsymbol{\Tilde{x}}$ be the estimate of $\boldsymbol{x}_0$ obtained by solving \eqref{eq:opt}. Then, a couple of questions are appropriate. How far is $\boldsymbol{\Tilde{x}}$ from $\boldsymbol{x}_0$? How does the estimation error scale with the number of correct correspondences $m$ and other variables such as $k$, $p$, and $d$? The answers to these questions are provided in the following section, where we present theoretical bounds on the error.

\section{Theoretical Guarantee}
Consider the optimization problem in \eqref{eq:opt}. An upper bound on the error $\|\boldsymbol{\Tilde{x}} - \boldsymbol{x}_0\|_2$ is presented in the following theorem:
\begin{theorem} Consider the optimization problem \eqref{eq:opt} and the observation model \eqref{obs_model} under assumptions (A1)-(A4). With $\lambda = 4(1+M) \sigma \sqrt{\frac{2\log{p}}{p}}$ for any $M \geq 0$, there exist constants $c_1, c_2, \varepsilon$ so that if $k \leq c_1 \frac{p-d}{\log{\frac{p}{k}}}$ and $\alpha \log{(p)} < (\sqrt{N} - \sqrt{d})$ for any $\alpha>0$, then the following inequality holds with probability at least $1-2\exp(-c_2(p-d)) - 2p^{-M^2}
-\exp{(-\log^2{(p)}/2)} - 2\exp{(-\alpha^2 \log^2{(p)}/2)}  - \exp{(-\alpha \log{(p)})}
$-

\begin{multline}\label{bound}
 \|\boldsymbol{\Tilde{x}} - \boldsymbol{x}_0\|_2 \leq \sigma \frac{\sqrt{d + 2\sqrt{d \alpha\log{p}} + 2\alpha\log{p}}}{\sqrt{m+p} - \sqrt{d} - \alpha\log{p}} +
 48(1+M)\sigma \varepsilon^{-1} \frac{(\sqrt{p} + \sqrt{d} + \log{p})}{(\sqrt{m+p} - \sqrt{d} - \alpha\log{p})^2} \frac{p}{p-d} \sqrt{k \log{p}}.
\end{multline}
\end{theorem}
The proof of the theorem follows the lines of proof in \cite{bendavid}, and the details are discussed in the Appendix. A few insights on the upper bound in \eqref{bound} are as follows.
\begin{enumerate}
    \item Similar to the bound obtained in \cite{bendavid}, the error term breaks into two components here as well. The first term is the error one would have incurred if the correspondences had been fully known $(k = 0)$. The second term is the excess error incurred for not knowing the correspondences.

    \item In the noiseless case $(\sigma = 0)$, perfect reconstruction of $\boldsymbol{z}$ is possible with high probability, and hence, the unknown vector $\boldsymbol{x}$ can be perfectly reconstructed with high probability.
    
    \item For a fixed number of measurements $p$, if we get more number of known correspondences $m$, then the error term falls off as $\frac{1}{(\sqrt{m+p}-\sqrt{d} - \alpha\log{p})}$ + $\frac{1}{(\sqrt{m+p}-\sqrt{d} - \alpha\log{p})^2}$. 
\end{enumerate}
In a nutshell, the bounds imply that the knowledge of correct correspondences helps improve the estimation accuracy, which is verified by simulation in the next section.


\section{Numerical Results}
To assess the proposed algorithm and compare it with robust regression, $\boldsymbol{x}_0$ was generated randomly and was kept fixed. Entries of $\mathbf{A}$ are sampled independently from $\mathcal{N}(0,1)$. For a given noise level, \eqref{eq:opt} is solved using CVXPY. For an objective comparison, we compute normalized reconstruction error  $\frac{\|\boldsymbol{\hat{x}} - \boldsymbol{x}_0\|_2}{\|\boldsymbol{x}_0\|_2}$, where $\boldsymbol{\hat{x}}$ is an estimate of $\boldsymbol{x}_0$ obtained by either the proposed procedure from \ref{eq:opt} or the robust regression method from \ref{robust_regression_estimator}. For a given permutation level, $k/p$, and noise level, the error is averaged over ten randomly generated permutation matrices and 50 independent noise realizations for each permutation matrix. The standard deviation $\sigma$ of noise is chosen as the specified noise percentage times the mean absolute value of the entries in the noiseless measurement vector $\boldsymbol{Ax}_0$.

\begin{figure}
\centering
\includegraphics[width= 5 in]{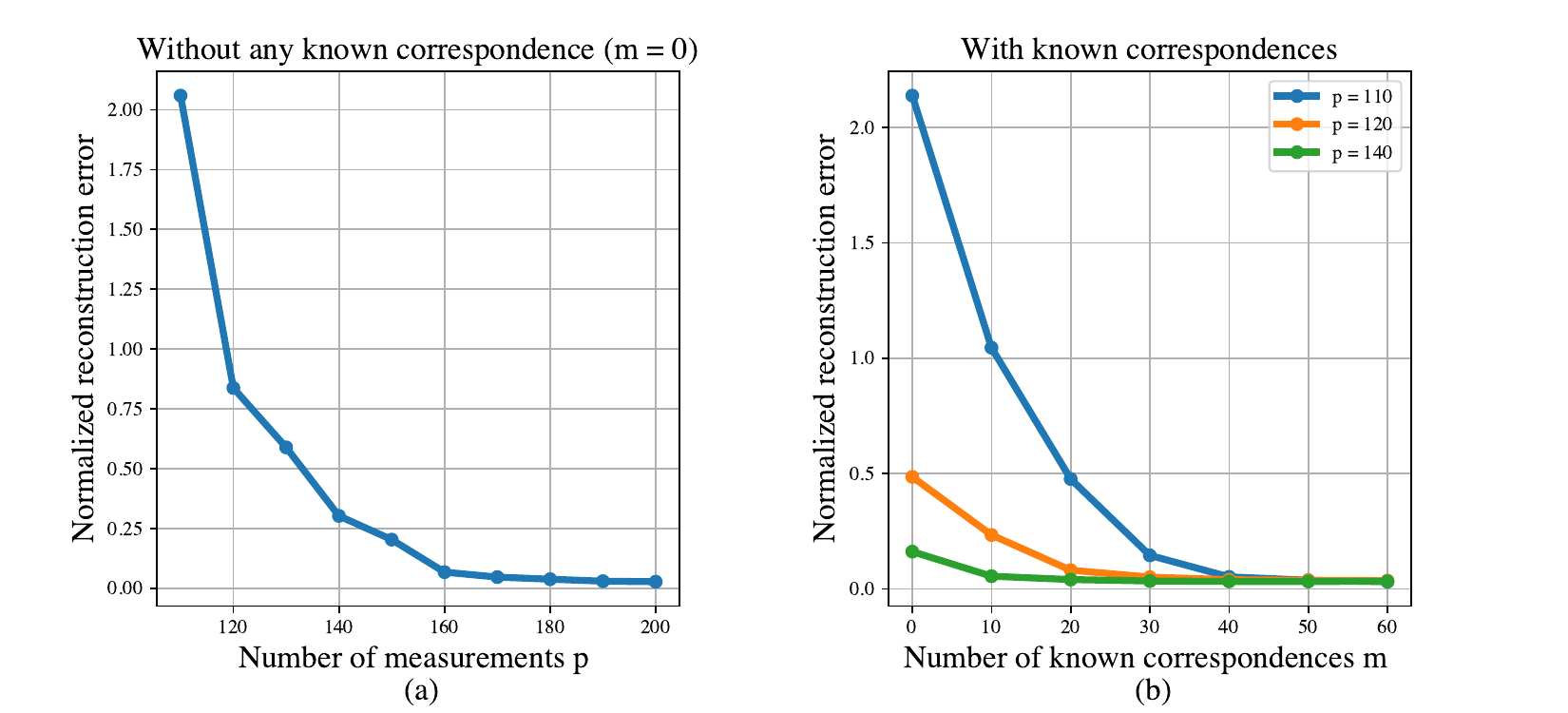}
\caption{Assessment of the effect of known correct correspondence on reconstruction error: (a) error as a function of $p$ when $m=0$. (b) error as a function of $m$ for different values of $p$. Known correspondence results in a lower error for a given $p$.}
\label{fig:errorVSpm}
\label{figure1}
\end{figure}
\begin{figure}[!t]
\centering
\includegraphics[width = 5 in]{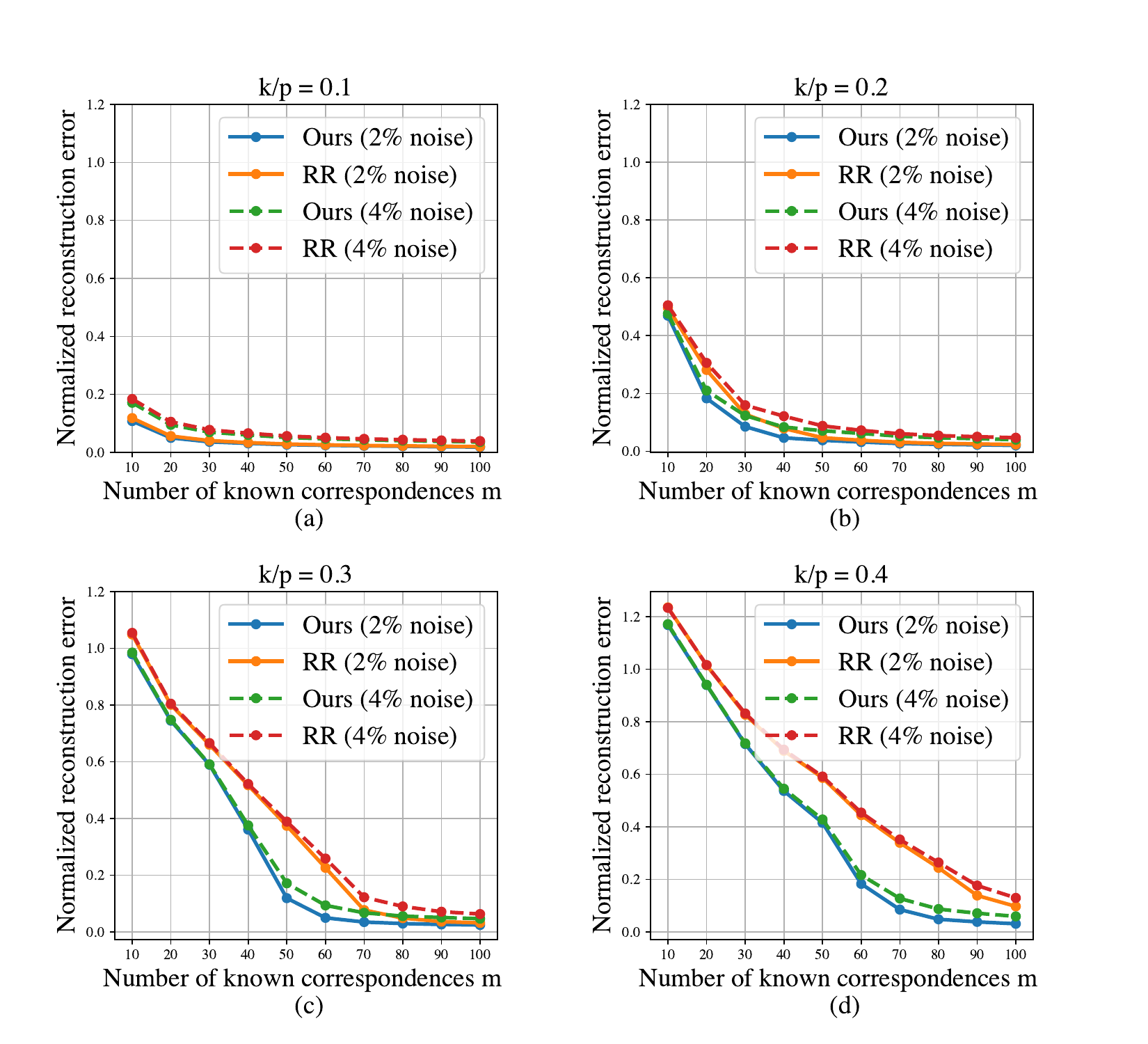}
\caption{A comparison of the proposed method and robust regression for $d = 100$, $p = 150$, and permutation level $k/p \in \{0.1, 0.2, 0.3, 0.4\}$: For low $k/p$, both methods perform equally well, however, for higher permutation levels, the proposed method results from \eqref{eq:opt} in lower error compared to robust regression from \eqref{robust_regression_estimator}. }
\label{fig:rr_vs_our_method_2_percent_noise}
\end{figure}

Our first objective is to assess the effect of the number of correct correspondences in reducing the estimation error. For this simulation we consider $d = 100$, $k/p = 0.1$ and $2\%$ measurement noise. In Fig.~\ref{fig:errorVSpm}(a), the error is plotted as a function of the number of measurements $p$ when there are no correct correspondences ($m= 0$). As expected, error reduces as $p$ increases. In Fig.~\ref{fig:errorVSpm}(b), we have shown errors as a function of $m$ for $p = 110, 120$, and $140$. Comparing errors in Fig.~\ref{fig:errorVSpm}(a) with those in Fig.~\ref{fig:errorVSpm}(b) for a given $p$, we note that having correct correspondences significantly reduces the error. For example, for $p = 140$, an addition of only $m =20$ measurements with known correspondences results in a reconstruction error of $0.04$, as compared to $0.16$ for $m = 0$. Alternatively, we infer that for a given error threshold, $p$ can be reduced by using a few correct correspondences. For example, the combinations $(p = 170, m=0)$ and $(p = 110, m=40)$ will result in $0.05 \%$ error.

Next, we compare the proposed method and its estimations with those of robust regression for $d = 100$ and $p = 150$. Errors for both the methods for $2\%$ and $4\%$ noise levels and different permutation levels $k/p$ are shown in Fig.~\ref{fig:rr_vs_our_method_2_percent_noise}. Though our methods always result in lower error than robust regression, the difference in performance varies greatly with the permutation level. For instance, at $k/p = 0.1$, the gain of the proposed method compared to robust regression is negligible. But it increases with $k/p$, that is, as the amount of permutation noise increases. For example, at $k/p = 0.4$, with $m = 80$ known correspondences, the robust regression estimator gives a reconstruction error of $0.24$, while that with our estimator is $0.05$ for $2\%$ noise.

After demonstrating through simulation that the suggested technique reduces errors and demands fewer measurements, our attention shifts to a subsequent application.

\begin{figure}[!t]
\centering
\includegraphics[width= 5 in]{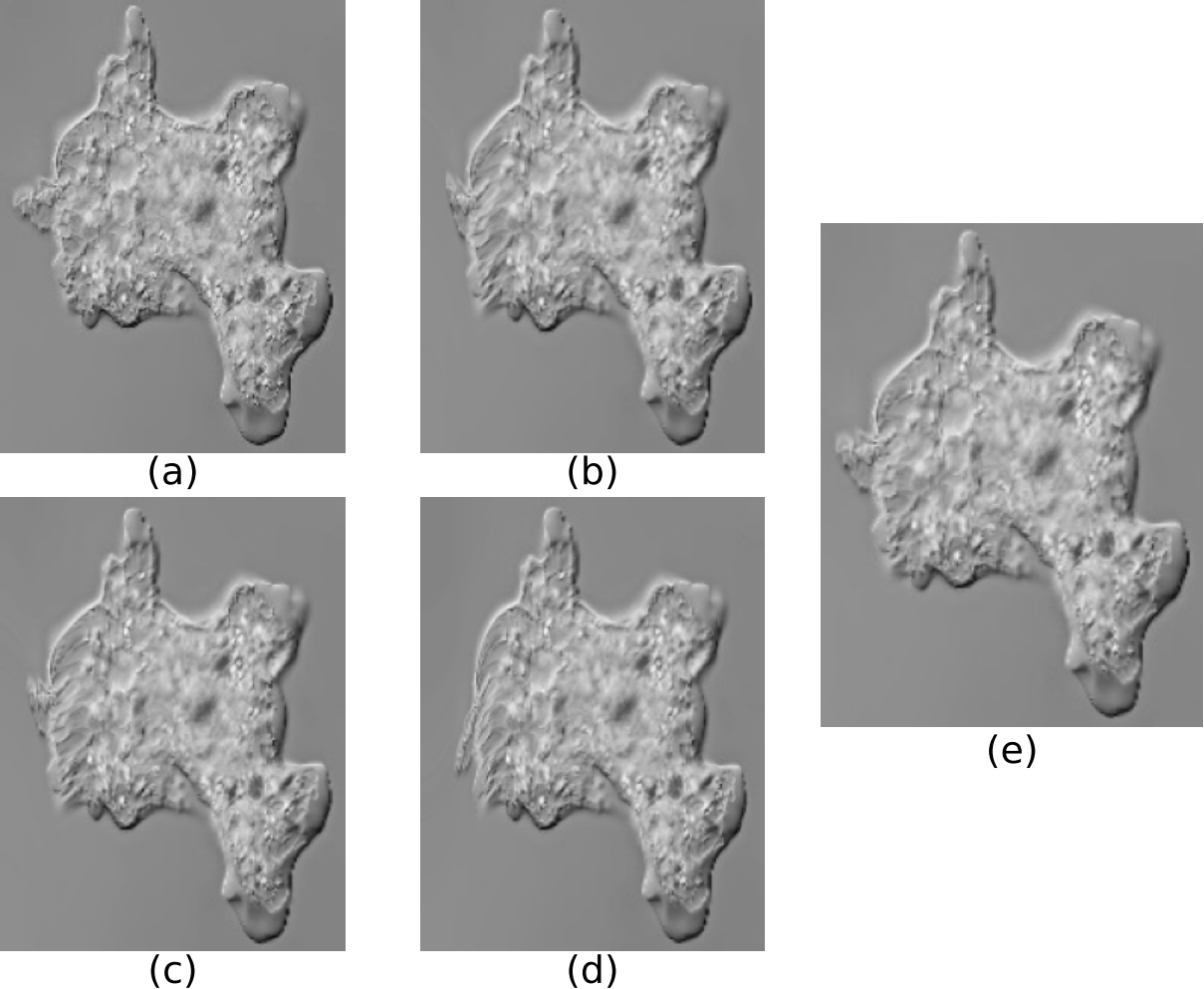}
\caption{(a): Base image $I$, (b): motion-deformed image $M$ using ground truth motion, reconstructed motion-deformed images using point-pairs from (c):  $S_1 \cup S_2$ (method \textsf{C3}), (d): only $S_1$ (method \textsf{C1}), (e): only $S_2$ (method \textsf{C2}). The reconstruction (c) looks visually more accurate.}
\label{fig:motion_deformation}
\end{figure}

\begin{figure}[!t]
\centering
\includegraphics[width= 5 in]{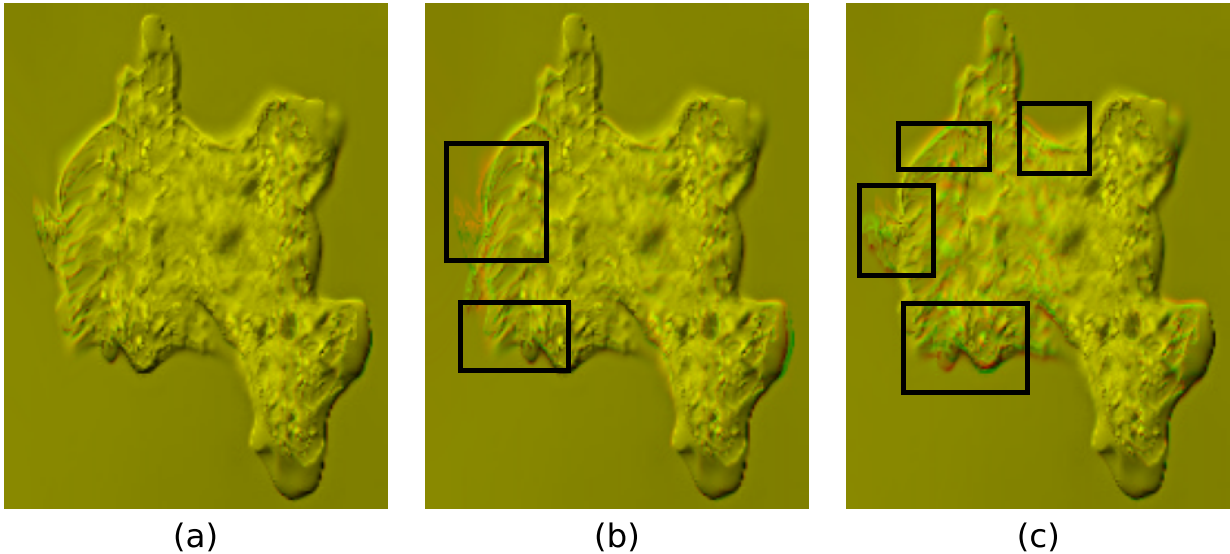}
\caption{Overlay of ground truth motion-deformed image (R channel) and motion-deformed image (G channel) using (a): method \textsf{C3}, (b): method \textsf{C2} and (c): method \textsf{C1}. Observe alignment using \textsf{C1}, \textsf{C2} is worse (many red or green edges, bordered by black boxes) than that with \textsf{C3}. The B channel of the overlay images is set to 0.}
\label{fig:overlays}
\end{figure}
\section{Application in Image Alignment}
The deformation between pairs of images in biomedical applications is often modeled as a non-rigid motion vector field, which is compactly expressed by a linear combination of some $d$ low-frequency 2D Discrete Cosine Transform (DCT) basis vectors \cite{motion_estimation}. Consider a reference image $I$ and a moving image $M$ which is a motion-deformed version of $I$, both of size $H \times W$. We define $\boldsymbol{u_1}, \boldsymbol{u_2} \in \mathbb{R}^{HW}$ as the vectorized displacement fields from $I$ to $M$ in the $X,Y$ directions respectively. Let $\boldsymbol{U} \in \mathbb{R}^{HW \times d}, d \ll HW$ denote the sub-matrix consisting of the first $d$ columns of the 2D DCT matrix of size $HW$ by $HW$. Then, we express $\boldsymbol{u_1} = \boldsymbol{U \theta_1}, \boldsymbol{u_2} = \boldsymbol{U \theta_2}$,
where $\boldsymbol{\theta_1}, \boldsymbol{\theta_2} \in \mathbb{R}^d$ are unknown 2D-DCT coefficient vectors. In some cases, we have displacement vector information at only a subset of pixels $S$ in $I$, as these can be obtained by salient feature point matching \cite{sift} or selected by domain experts. Then we have $\boldsymbol{u_1}\big|_S = \boldsymbol{U}\big|_S \boldsymbol{\theta_1}, \boldsymbol{u_2}\big|_S = \boldsymbol{U}\big|_S \boldsymbol{\theta_2}$, where $\boldsymbol{U}\big|_S \in \mathbb{R}^{|S| \times d}$ contains the rows from $\boldsymbol{U}$ corresponding to the pixel locations in $S$ and $\boldsymbol{u_1}\big|_S, \boldsymbol{u_2}\big|_S \in \mathbb{R}^{|S|}$ are sub-vectors of $\boldsymbol{u_1}, \boldsymbol{u_2}$ respectively containing only vectors from locations in $S$. The goal is to estimate $\boldsymbol{\theta_1}, \boldsymbol{\theta_2}$ given $S$, $\boldsymbol{u_1}\big|_S$, $\boldsymbol{u_2}\big|_S$. To this end, we use SIFT descriptors \cite{sift} to obtain a set $S_1$ of $p$ key-point pairs in the two images. Further, we accurately annotate a set of $m$ corresponding point-pairs in the images $I$ and $M$, which we refer to as $S_2$. The correspondences of the $m$ point-pairs in $S_2$ are known \emph{accurately}, whereas the correspondences in a \emph{small} number of the $p$ point-pairs in $S_1$ may be incorrect due to errors in SIFT-based point matching. These errors can be modeled as sparse permutations. Note that the indices of the point-pairs in the erroneous subset of $S_1$ are also unknown. In the presence of the underlying permutation noise, we can write the modified observation model as  
$\boldsymbol{u_1}\big|_{S_1\cup S_2} = \boldsymbol{P_1 U}\big|_{S_1\cup S_2} \boldsymbol{\theta_1}, \boldsymbol{u_2}\big|_{S_1\cup S_2} = \boldsymbol{P_2 U}\big|_{S_1\cup S_2} \boldsymbol{\theta_2}$, where $\boldsymbol{P_1, P_2}$ are unknown permutation matrices. Note that the correspondences of $m$ measurements are known, while the correspondences of the remaining $p$ measurements may contains a small number of errors, and hence the aforementioned model is similar to \eqref{obs_model}. Therefore, we can use our framework from \eqref{eq:opt} to estimate $\boldsymbol{\theta_1}, \boldsymbol{\theta_2}$ given $S_1, S_2$, $\boldsymbol{u_1}\big|_{S_1\cup S_2}, \boldsymbol{u_2}\big|_{S_1\cup S_2}$. In this experiment, we set $d = 10$ and synthetically generate motion using  $\boldsymbol{u_1} = \boldsymbol{U\theta_1}, \boldsymbol{u_2} = \boldsymbol{U\theta_2}$. We use the SIFT descriptor technique to obtain a set of $p = 179$ key-point pairs in the two images, which form the set $S_1$. Further, we accurately annotate a set of $m = 8$ point-pairs, which form $S_2$. Note that we have kept $m < d$ to avoid the trivial scenario, where reconstruction can be done only using $S_2$, and also because manual annotation of a larger number of point-pairs is often not feasible. We reconstruct the motion-deformed image by estimating $\boldsymbol{\theta_1}, \boldsymbol{\theta_2}$, and thus $\boldsymbol{u_1}, \boldsymbol{u_2}$, and then applying this motion to the reference image $I$. We estimate $\boldsymbol{\theta_1}, \boldsymbol{\theta_2}$ in 3 ways via the model from \eqref{eq:opt}: (\textsf{C1}) using only the point-pairs from $S_1$, (\textsf{C2}) using only the manually annotated point-pairs from $S_2$, and (\textsf{C3}) using point-pairs from $S_1 \cup S_2$. The reference image $I$, the motion-deformed image $M$ using ground truth motion, as well as the motion-deformed images using motion obtained via \textsf{C1}, \textsf{C2}, \textsf{C3}, are plotted in Fig.~\ref{fig:motion_deformation}. 
The normalized mean squared error (NMSE) between the original motion-deformed image and the reconstructed motion-deformed image for \textsf{C1}, \textsf{C2}, \textsf{C3} are respectively 0.008, 0.005 and 0.002, showing the superior performance of \textsf{C3}. Also, observe the overlay images of the ground-truth motion-deformed image and the motion-deformed image using the motion estimates from methods \textsf{C3}, \textsf{C1} and \textsf{C2}. These are plotted in Fig.~\ref{fig:overlays}. The overlay for \textsf{C3} shows significantly fewer red or green edges as compared to the other two. 
\begin{figure}[!t]
\centering
\includegraphics[width=\columnwidth]{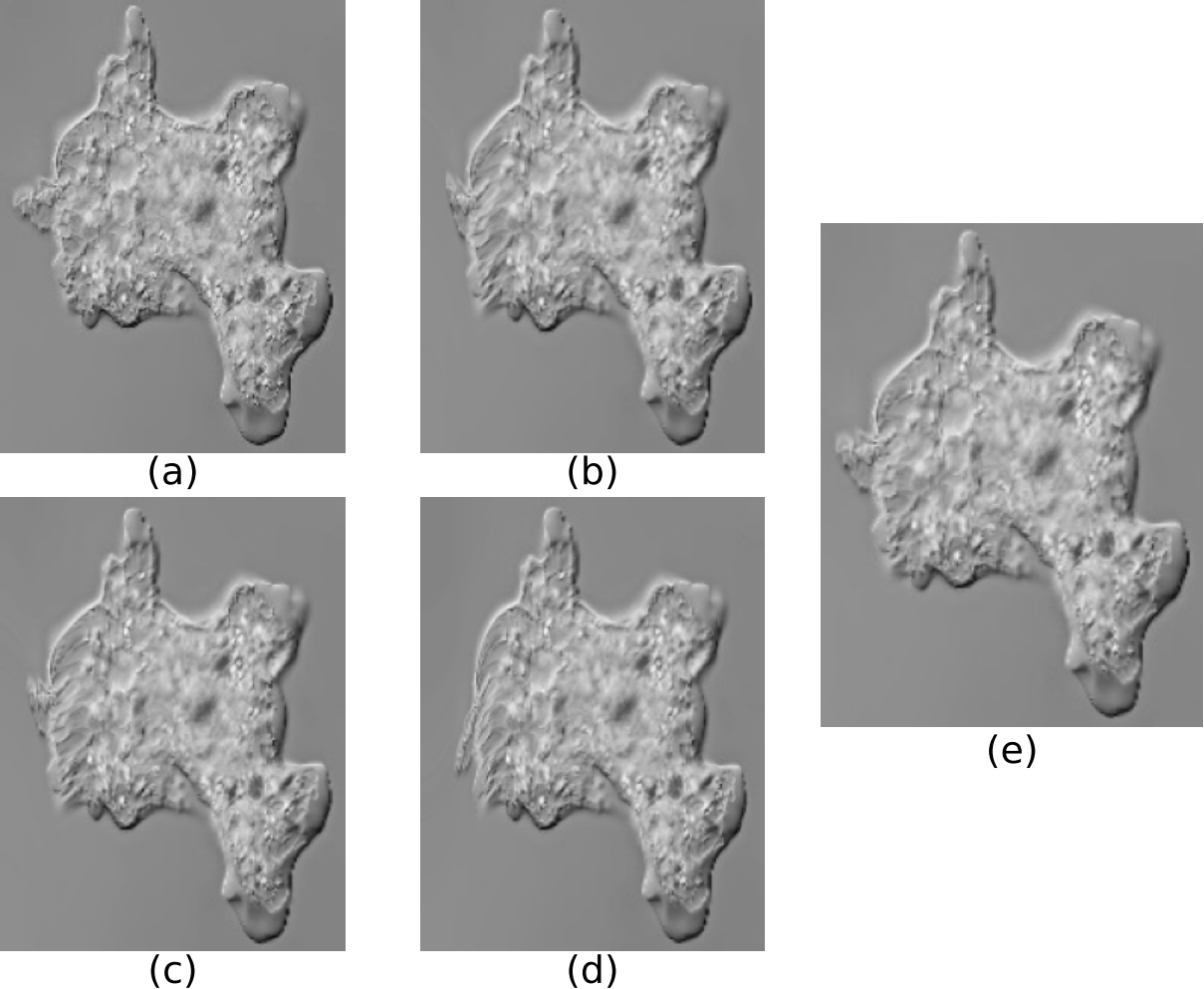}
\caption{(a): Base image $I$, (b): motion-deformed image $M$ using ground truth motion, reconstructed motion-deformed images using point-pairs from (c):  $S_1 \cup S_2$ (method \textsf{C3}), (d): only $S_1$ (method \textsf{C1}), (e): only $S_2$ (method \textsf{C2}). The reconstruction (c) looks visually more accurate.}
\label{fig:motion_deformation}
\end{figure}

\begin{figure}[!t]
\centering
\includegraphics[width=\columnwidth]{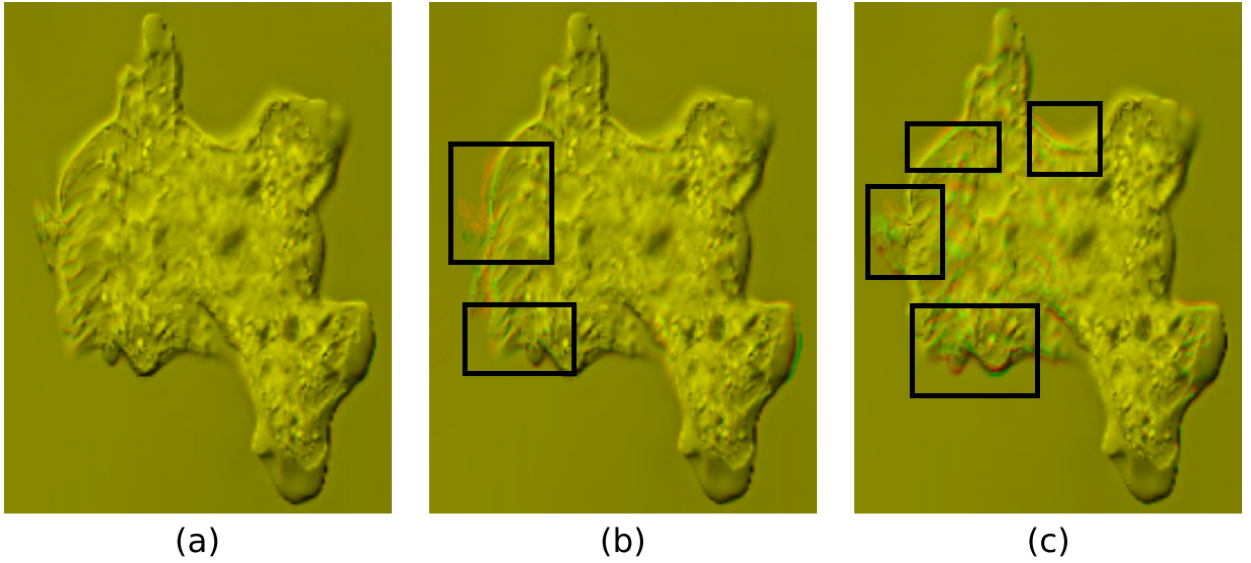}
\caption{Overlay of ground truth motion-deformed image (R channel) and motion-deformed image (G channel) using (a): method \textsf{C3}, (b): method \textsf{C2} and (c): method \textsf{C1}. Observe alignment using \textsf{C1}, \textsf{C2} is worse (many red or green edges, bordered by black boxes) than that with \textsf{C3}. The B channel of the overlay images is set to 0.}
\label{fig:overlays}
\end{figure}

\section{Conclusion}
We proposed an algorithm to estimate the unknown vector in unlabelled sensing with sparse permutations given a small number of measurements with known correspondences. We derived a theoretical upper bound on the reconstruction error. Through simulations, we showed that a few measurements with known correspondences can significantly improve the reconstruction error, or reduce the sample complexity for the same reconstruction error as obtained without known correspondences. We found several regimes where our estimator significantly outperforms robust regression techniques while maintaining an acceptable level of reconstruction error. Lastly, we consider an application in DCT-based motion estimation. We showed that a few manually annotated point-pairs with accurate correspondence, along with the SIFT key point-pairs (where some correspondences can be erroneous), can improve motion estimation.

\section*{Appendix}
\begin{proof}

To prove Theorem~1, the cost function to be minimized  
is given as
\begin{equation}
    L(\boldsymbol{x, z}) = \|\boldsymbol{y_1-A_1 x}\|_2^2+\|\boldsymbol{y_2-A_2 x-z}\|_2^2 + \lambda_1 \|\boldsymbol{z}\|_1. 
\end{equation}
Next, we have that
\begin{equation}
    \begin{split}
\boldsymbol{\Tilde{x}}, \boldsymbol{\Tilde{z}} &= \underset{\boldsymbol{x} \in \mathbb{R}^d, \boldsymbol{z} \in \mathbb{R}^p} {\operatorname{arg} \min} L(\boldsymbol{x, z}) \\
&= \underset{\boldsymbol{x} \in \mathbb{R}^d, \boldsymbol{z} \in \mathbb{R}^p} {\operatorname{arg} \min} \|\boldsymbol{y_1-A_1 x}\|_2^2+\|\boldsymbol{y_2-A_2 x-z}\|_2^2 + \lambda_1 \|\boldsymbol{z}\|_1.         
    \end{split}
\end{equation}
We perform a reparameterization $\boldsymbol{e} = \boldsymbol{z}/\sqrt{p}$ and write the above equation as
\begin{equation}\label{eq1}
  \boldsymbol{\Tilde{x}}, \boldsymbol{\Tilde{e}} = \underset{\boldsymbol{x} \in \mathbb{R}^d, \boldsymbol{e} \in \mathbb{R}^p} {\operatorname{arg} \min } \frac{1}{p}\|\boldsymbol{y_1-A_1 x}\|_2^2+\frac{1}{p}\|\boldsymbol{y_2-A_2 x} - \sqrt{p} \boldsymbol{e}\|_2^2 + \lambda\|\boldsymbol{e}\|_1,  
\end{equation}where $\lambda = \frac{\lambda_1}{\sqrt{p}} > 0$.
The above optimization involves minimizing over two variables. To simplify things, we project the vector $\boldsymbol{y_2-A_2 x} - \sqrt{p} \boldsymbol{e}$, which is a function of both $\boldsymbol{x}$ and $\boldsymbol{e}$, into two components by projecting it onto the column space of $\boldsymbol{A}_2$ and its orthogonal complement. Given that the entries of $\boldsymbol{A_2}$ are i.i.d. zero mean unit variance Gaussian random variables,  $\boldsymbol{A_2}$ is full-column rank with probability 1. The projection matrix $\boldsymbol{H}$ which projects onto the column space of $\boldsymbol{A_2}$ is given as $\boldsymbol{H} = \boldsymbol{A_2(A_2^T A_2)^{-1} A_2^T}$. Let $\boldsymbol{H^{\perp}}$ denote a projection matrix which projects onto the orthogonal complement of the column space of $\boldsymbol{A_2}$. Then, by using the decomposition $\boldsymbol{y_2-A_2 x} - \sqrt{p} \boldsymbol{e} = \boldsymbol{H} \left( \boldsymbol{y_2-A_2 x} - \sqrt{p} \boldsymbol{e} \right) + \boldsymbol{H^{\perp}} \left( \boldsymbol{y_2-A_2 x} - \sqrt{p} \boldsymbol{e} \right)$ and noting that the two terms are orthogonal we can rewrite \eqref{eq1} as
    \begin{equation} \label{eq11}
   \boldsymbol{\Tilde{x}}, \boldsymbol{\Tilde{e}} = \underset{\boldsymbol{x} \in \mathbb{R}^d, \boldsymbol{e} \in \mathbb{R}^p} {\operatorname{arg} \min } \frac{1}{p}\|\boldsymbol{y_1-A_1 x}\|_2^2+\frac{1}{p}\|\boldsymbol{H(y_2}-\sqrt{p} \boldsymbol{e}) - \boldsymbol{A_2 x}\|_2^2 +  \frac{1}{p}\|\boldsymbol{H^{\perp}(y_2}-\sqrt{p} \boldsymbol{e})\|_2^2+ \lambda\|\boldsymbol{e}\|_1.
 \end{equation}
With this revised formulation, we note that the term  $G(\boldsymbol{e}) := \frac{1}{p}\|\boldsymbol{H^{\perp}(y_2}-\sqrt{p} \boldsymbol{e})\|_2^2+ \lambda\|\boldsymbol{e}\|_1$ in \eqref{eq11} is independent of $\boldsymbol{x}$. We solve \eqref{eq1} or \eqref{eq11} by first optimizing $G(\boldsymbol{e})$ and then plugging the solution into the rest of the terms $\frac{1}{p}\|\boldsymbol{y_1-A_1 x}\|_2^2+\frac{1}{p}\|\boldsymbol{H(y_2}-\sqrt{p} \boldsymbol{e}) - \boldsymbol{A_2 x}\|_2^2$ to find an optimal estimate of $\boldsymbol{x}$. Specifically, we first estimate 
\begin{equation}
    \boldsymbol{\Tilde{e}} = \underset{\boldsymbol{e} \in \mathbb{R}^p} {\operatorname{arg} \min } \frac{1}{p}\|\boldsymbol{H^{\perp}(y_2}-\sqrt{p} \boldsymbol{e})\|_2^2+ \lambda\|\boldsymbol{e}\|_1.
\end{equation}
Then, we use $\boldsymbol{\Tilde{e}}$ to get a closed form expression for $\boldsymbol{\Tilde{x}}$ as
\begin{equation}
\begin{split}
    \boldsymbol{\Tilde{x}} = \underset{\boldsymbol{x} \in \mathbb{R}^d} {\operatorname{arg} \min } \left\|\boldsymbol{\begin{bmatrix}
        \boldsymbol{A_1} \\
        \boldsymbol{A_2}
    \end{bmatrix} x}  - \begin{bmatrix}
        \boldsymbol{y_1} \\
        \boldsymbol{H(y_2}-\sqrt{p} \boldsymbol{\Tilde{e}}) 
    \end{bmatrix}\right\|_2^2 
    =  \underset{\boldsymbol{x} \in \mathbb{R}^d} {\operatorname{arg} \min } \|\boldsymbol{Ax-h}\|_2^2
    = \boldsymbol{A^\dagger h}, 
\end{split}
\end{equation}
where $\boldsymbol{A^\dagger}$ denotes the Moore-Penrose pseudo-inverse of $\boldsymbol{A}$ and $\boldsymbol{h} := \begin{bmatrix}
        \boldsymbol{y_1} \\
        \boldsymbol{H(y_2}-\sqrt{p} \boldsymbol{\Tilde{e}}) 
    \end{bmatrix}$. 
Next, by using the estimated $\boldsymbol{\Tilde{x}}$, we derive bounds on the error $\|\boldsymbol{\Tilde{x}} - \boldsymbol{x_0}\|_2$. Starting from
\begin{equation}
    \boldsymbol{\Tilde{x}} = \boldsymbol{A^\dagger h} = \boldsymbol{(A^T A)^{-1}A^T h},
\end{equation}    
we use the following steps.
\begin{equation}
    \begin{split}
        \boldsymbol{A^T A \Tilde{x}} &= \boldsymbol{A^T h} \\
                                &= \begin{bmatrix}
                            \boldsymbol{A_1^T} & \boldsymbol{A_2^T}
                        \end{bmatrix} \begin{bmatrix}
                            \boldsymbol{y_1} \\
                            \boldsymbol{H(y_2}-\sqrt{p} \boldsymbol{\Tilde{e}})
                        \end{bmatrix}\\
                        &= \boldsymbol{A_1^T y_1} + \boldsymbol{A_2^T H (y_2} -\sqrt{p}\boldsymbol{\Tilde{e}})\\
                        &= \boldsymbol{A_1^T A_1 x_0 + A_1^T \epsilon_1 + A_2^T(y_2} - \sqrt{p}\boldsymbol{\Tilde{e}} + \sqrt{p}\boldsymbol{e_0} - \sqrt{p}\boldsymbol{e_0})\\
                        &= \boldsymbol{A_1^T A_1 x_0} + \boldsymbol{A_1^T \epsilon_1} + \boldsymbol{A_2^T}(\boldsymbol{y_2} - \sqrt{p}\boldsymbol{e_0}) + \sqrt{p}\boldsymbol{A_2^T} (\boldsymbol{e_0} -\boldsymbol{\Tilde{e}})\\
                        &= \boldsymbol{A_1^T A_1 x_0} + \boldsymbol{A_1^T \epsilon_1} + \boldsymbol{A_2^T}(\boldsymbol{A_2 x_0} + \boldsymbol{\epsilon_2}) + \sqrt{p}\boldsymbol{A_2^T} (\boldsymbol{e_0} -\boldsymbol{\Tilde{e}})\\
                        &= \boldsymbol{A^T A x_0} + \boldsymbol{A^T \epsilon} + \sqrt{p}\boldsymbol{A_2^T} (\boldsymbol{e_0} -\boldsymbol{\Tilde{e}}) 
    \end{split}
\end{equation}
or
\begin{equation}
 \boldsymbol{A^T A (\Tilde{x} - x_0)} = \boldsymbol{A^T \epsilon} + \sqrt{p}\boldsymbol{A_2^T} (\boldsymbol{e_0} -\boldsymbol{\Tilde{e}})   
\end{equation}
or 
\begin{equation}
\boldsymbol{\Tilde{x} - x_0} = \boldsymbol{A^\dagger \epsilon} + \sqrt{p}\boldsymbol{(A^T A)^{-1} A_2^T } (\boldsymbol{e_0} -\boldsymbol{\Tilde{e}}).
\end{equation}
We upper-bound the quantity on the right using the standard norm inequalities. 
\begin{equation}\label{eq2}
    \begin{split}
        \|\boldsymbol{\Tilde{x}} - \boldsymbol{x_0}\|_2 & \leq  \|\boldsymbol{A^\dagger \epsilon}\|_2 + \sqrt{p} \|\boldsymbol{(A^T A)^{-1} A_2^T }\|_2 \|\boldsymbol{e_0} -\boldsymbol{\Tilde{e}}\|_2\\
        & \leq \|\boldsymbol{A^\dagger \epsilon}\|_2 + \sqrt{p} \frac{\|\boldsymbol{A_2}\|_2}{(\sigma_{\text{min}}(\boldsymbol{A}))^2} \|\boldsymbol{e_0} -\boldsymbol{\Tilde{e}}\|_2.
    \end{split}
\end{equation}

In order to obtain a bound in terms of $m, p, d, k$ and other parameters, we use a series of concentration inequalities to bound the quantities on the RHS of \eqref{eq2}. To this end, we use Lemmas 1-4 provided after the proof. From Lemma \ref{lemma1}, using $\boldsymbol{A_2}$ and $\boldsymbol{A}$ in the role of $\boldsymbol{X}$, for any $t_1, t_2 > 0$ such that $t_2 < \sqrt{N} - \sqrt{d}$, we have that

\begin{equation}\label{eq3}
    \mathbb{P}(\|\boldsymbol{A_2}\|_2 \leq \sqrt{p} + \sqrt{d} + t_1) \geq 1 - \exp(-t_1^2/2),
\end{equation}
and
\begin{equation}\label{eq4}
   \mathbb{P}\bigg(\frac{1}{\sigma_{\text{min}}(\boldsymbol{A})} \leq \frac{1}{\sqrt{N} - \sqrt{d} - t_2}\bigg) \geq 1 - 2 \exp(-t_2^2/2). 
\end{equation}
From Lemma \ref{lemma3}, using $\boldsymbol{A}$ in the role of $\boldsymbol{X}$ and $\boldsymbol{\epsilon}$ in the role of $\boldsymbol{g}$ with $t = t_2$, we have
\begin{equation}\label{eq5}
    \mathbb{P}\bigg(\|\boldsymbol{A^\dagger \epsilon}\|_2 \leq \sigma \frac{\sqrt{d + 2\sqrt{t_2d} + 2t_2}}{\sigma_{\text{min}}(\boldsymbol{A})} \bigg) \geq 1 - \exp(-t_2).
\end{equation}
We use Lemma \ref{lemma4} to upper bound $\|\boldsymbol{e_0}-\boldsymbol{\tilde{e}}\|_2$ and the concentration inequalities \eqref{eq3}, \eqref{eq4}, \eqref{eq5} with $t_1 = \log{p}, t_2 = \alpha \log{p}$ where $\alpha \in \mathbb{R} > 0$ to upper bound $\|\boldsymbol{A^{\dagger}\epsilon}\|_2$, in \eqref{eq2} to conclude the proof, producing the following final bound:
\begin{equation}
 \|\boldsymbol{\Tilde{x}} - \boldsymbol{x}_0\|_2 \leq \sigma \frac{\sqrt{d + 2\sqrt{d \alpha\log{p}} + 2\alpha\log{p}}}{\sqrt{m+p} - \sqrt{d} - \alpha\log{p}} +
 48(1+M)\sigma \varepsilon^{-1} \frac{(\sqrt{p} + \sqrt{d} + \log{p})}{(\sqrt{m+p} - \sqrt{d} - \alpha\log{p})^2} \frac{p}{p-d} \sqrt{k \log{p}}.    
\end{equation}

\end{proof}


\begin{lemma}[\cite{concentration_inequalities}]\label{lemma1}
Let $\boldsymbol{X}$ be an $m \times n$ Gaussian random matrix with i.i.d. $\mathcal{N}(0, 1)$ entries. Then, for any $t>0$, we have

\[
\mathbb{P}(\|\boldsymbol{X}\|_2 \geq \sqrt{m} + \sqrt{n} + t) \leq \exp(-t^2/2)
\]
and
\[
\mathbb{P}(\sigma_{\text{min}}(\boldsymbol{X}) \geq \sqrt{m} - \sqrt{n} - t) \geq 1 - 2 \exp(-t^2/2).
\]

\end{lemma}

\begin{lemma}[\cite{MR2994877}]\label{lemma2}
    Let $\boldsymbol{Z}$ be an $m\times n$ matrix, define $\boldsymbol{\Gamma} := \boldsymbol{Z^TZ}$ and $\boldsymbol{g} \sim \mathcal{N}(0, \sigma^2 I_n)$. Then, for any $t>0$, we have
    \[
\mathbb{P}(\|\boldsymbol{Zg}\|_2^2 > \sigma^2(\text{tr}(\boldsymbol{\Gamma}) + 2\sqrt{\text{tr}(\boldsymbol{\Gamma}^2)t} + 2 \|\boldsymbol{\Gamma}\|_2t)) \leq \exp{(-t)}.  
    \]  
\end{lemma}

\begin{lemma}\label{lemma3}
    Let $\boldsymbol{X}$ be an $m \times n$ matrix and $\boldsymbol{g} \sim \mathcal{N}(0, \sigma^2 I_n)$. Then, for any $t>0$, we have 

    \[
    1 - \mathbb{P}\bigg(\|\boldsymbol{X^\dagger g}\|_2^2 \leq \sigma^2\bigg( \frac{n}{\sigma^2_{\text{min}}(\boldsymbol{X})} + \frac{2 \sqrt{n t}}{\sigma^2_{\text{min}}(\boldsymbol{X})} + \frac{2 t}{\sigma^2_{\text{min}}(\boldsymbol{X})}\bigg)\bigg) \leq \exp(-t).
    \]
\end{lemma}

\begin{proof}
    We use Lemma \ref{lemma2} with $\boldsymbol{X^\dagger}$ in the role of $\boldsymbol{Z}$. Then, $\boldsymbol{\Gamma} = \boldsymbol{X (X^T X)^{-2} X^T}$. We use the three results given in Appendix G in \cite{bendavid} to get
    \begin{enumerate}
        \item $\text{tr}(\boldsymbol{\Gamma}) \leq \frac{n}{\sigma^2_\text{min}{\boldsymbol{(X)}}}$,

        \item $\|\boldsymbol{\Gamma}\|_2 = \frac{1}{\sigma^2_\text{min}{\boldsymbol{(X)}}}$,

        \item $\sqrt{\text{tr}(\boldsymbol{\Gamma^2})} \leq \frac{\sqrt{n}}{\sigma^2_\text{min}{\boldsymbol{(X)}}}$.
    \end{enumerate} which are then plugged back in Lemma \ref{lemma2} to obtain the statement of the current lemma.   
\end{proof}

\begin{lemma}[See Lemma 4 and Lemma 6 in\cite{bendavid}]\label{lemma4}
    Let $\boldsymbol{A_2}$ be a $p\times d$ Gaussian random matrix with i.i.d. $\mathcal{N}(0, 1)$ entries. We have observation model as $\boldsymbol{y}_2 = \boldsymbol{P}_2\boldsymbol{A}_2 \boldsymbol{x}_0 + \boldsymbol{\epsilon}_2$ where $\boldsymbol{x_0} \in \mathbb{R}^{d}$ is unknown, $\boldsymbol{P_2} \in \mathbb{R}^{p\times p}$ is a $k$-sparse permutation matrix and $\boldsymbol{\epsilon_2} \in \mathbb{R}^{p}$ is noise vector with i.i.d $\mathcal{N}(0, \sigma^2)$ entries. We define the permutation error vector as $\boldsymbol{z}_0 = \boldsymbol{P}_2 \boldsymbol{A}_2 \boldsymbol{x}_0 - \boldsymbol{A}_2 \boldsymbol{x}_0 = \sqrt{p}\boldsymbol{e}_0$. Define $\boldsymbol{\Tilde{e}} := \underset{\boldsymbol{e} \in \mathbb{R}^p} {\operatorname{arg} \min } \frac{1}{p}\|\boldsymbol{H^{\perp}(y_2}-\sqrt{p} \boldsymbol{e})\|_2^2+ \lambda\|\boldsymbol{e}\|_1$ where $\boldsymbol{H^{\perp}}$ denotes a projection matrix which projects onto the orthogonal complement of the column space of $\boldsymbol{A_2}$. If $\lambda = 4(1+M) \sigma \sqrt{\frac{2\log{p}}{p}}$ for any $M \geq 0$, there exist constants $c_1, c_2, \varepsilon$ so that if $k \leq c_1 \frac{p-d}{\log{\frac{p}{k}}}$, then the following inequality holds with probability at least $1-2\exp(-c_2(p-d)) - 2p^{-M^2}$:
\[
    \|\boldsymbol{\Tilde{e} - e_0}\|_2 \leq 48(1+M)\sigma\frac{p}{p-d}\varepsilon^{-1} \sqrt{\frac{k\log{p}}{p}}.
\]
\end{lemma}

\bibliographystyle{IEEEtran} 
\bibliography{cites,US_refs}
                       
\end{document}